\documentclass[twocolumn,english,groupedaddress,superscriptaddress,aps,prx,10pt,floatfix]{revtex4-2}

\usepackage[T1]{fontenc}
\usepackage[utf8]{inputenc}

\usepackage{amsmath,amsthm,mathtools,amssymb}
\usepackage{graphicx}
\usepackage{dsfont}
\usepackage{url}

\usepackage[breaklinks]{hyperref}
\usepackage{todonotes}
\usepackage{orcidlink}
\usepackage{siunitx}
\usepackage{array}
\usepackage{float}
\usepackage{multirow}

\usepackage{bm}
\newcommand{\matr}[1]{\bm{\mathrm{#1}}}
\renewcommand{\vec}[1]{\bm{\mathrm{#1}}}

\newtheorem{alg}{Algorithm}

\newcommand{\eye}{\mathds{1}}

\begin{document}

\title{Phase locking and multistability in the topological Kuramoto model on cell complexes}

\author{Iva Ba\v{c}i\'{c}  \orcidlink{0000-0003-2987-5065}}
\thanks{iva@ipb.ac.rs, d.witthaut@fz-juelich.de}
\affiliation{Institute of Climate and Energy Systems: Energy Systems Engineering (ICE-1), Forschungszentrum J\"ulich, 52428 J\"ulich, Germany}
\affiliation{Institute of Physics Belgrade, University of Belgrade, Serbia}

\author{Michael T. Schaub
\orcidlink{0000-0003-2426-6404}}
\affiliation{RWTH Aachen University, Aachen, Germany}

\author{J\"urgen Kurths
\orcidlink{0000-0002-5926-4276}}
\affiliation{Potsdam Institute for Climate Impact Research (PIK), Potsdam, Germany}

\author{Dirk Witthaut \orcidlink{0000-0002-3623-5341}}
\thanks{iva@ipb.ac.rs, d.witthaut@fz-juelich.de}
\affiliation{Institute of Climate and Energy Systems: Energy Systems Engineering (ICE-1), Forschungszentrum J\"ulich, 52428 J\"ulich, Germany}
\affiliation{Institute for Theoretical Physics, University of Cologne, 50937 K\"oln, Germany}

\begin{abstract}
Higher-order interactions fundamentally shape collective dynamics in oscillator networks. The topological Kuramoto model captures these effects by extending synchronization models to include interactions between cells of arbitrary dimension within simplicial and cell complexes. We introduce the topological nonlinear Kirchhoff conditions to characterize all phase-locked states of the topological Kuramoto model. These states are organized by winding numbers associated with generalized independent cycles, which quantify how phases wind around these cycles. 
Using rings, Platonic solids, and regular simplices as illustrative examples, we uncover a universal rule: boundaries must have at least five elements for multistability to arise. We further find that independent winding numbers associated with lower- and higher-dimensional boundaries generate cascades of multistability across dimensions.
These results show how the topology and boundary structure of cell complexes influence phase locking and multistability, and provide a general framework for collective dynamics on cell complexes.
\end{abstract}

\maketitle

\section{Introduction}

Synchronization phenomena are a hallmark of many natural and engineered systems, from neuronal activity to power grid operation~\cite{pikovsky2001synchronization,dorfler2014synchronization}. 
The Kuramoto model of coupled oscillators has long served as a prototype for understanding how synchronization emerges from local interactions~\cite{kuramoto1975self,acebron2005kuramoto}. 
While early studies focused on globally coupled  systems~\cite{strogatz2000kuramoto}, later extensions to complex networks revealed that network topology crucially shapes both the onset and the nature of synchronization ~\cite{rodrigues2016kuramoto}. 
Such systems often exhibit multistability, where multiple distinct synchronized states coexist~\cite{manik2017cycle, delabays2016multistability,jafarpour2022flow, hellmann2020network}.

In the classical Kuramoto model, the coupling between the oscillators is pairwise. 
This assumption is valid for many systems, but in certain contexts where group interactions play an essential role, such as neuronal assemblies~\cite{Breakspear2010}, the brain connectome~\cite{Sizemore2018, Giusti2016, Andjelkovic2020}, multimode lasers~\cite{Nixon2013}, or protein interaction networks~\cite{Estrada2018}, it is too restrictive to capture all the relevant couplings in the system. 
Higher-order connectivity structures that encode such interactions are increasingly recognized as being crucial to many collective dynamics~\cite{Millan2025higherorderdynamics, battiston2020networks,battiston2021physics, Bianconi2021book, mulder2018network, Bick2023, Iacopini2019, Santoro2024, Zhang2023HigherOrder}. 
Recent advances in network science~\cite{battiston2020networks,battiston2021physics,Horak2009, Sizemore2018, Petri2013, Petri2014, Giusti2016, Wu2015, Iacopini2019, Andjelkovic2016, Andjelkovic2020, Calmon2022Localdirac, Calmon2022dirac, Calmon2023dirac} 
have introduced models where the dynamics is not defined only on nodes, but also lives on edges, faces, or higher-dimensional elements of simplicial or cell complexes. Cell complexes are higher-order network structures built from cells of different dimensions, such as nodes, edges, polygons, polytopes, and higher-dimensional analogues~\cite{hoppe2025don, Sizemore2018,Roddenberry2022}. 
Most studies on synchronization focus on the special case of simplicial complexes~\cite{millan2020explosive,Nurisso2024UnifiedKuramoto}, where each cell is a simplex. However, many real-world systems, such as social interactions or protein complexes, are better described by cell complexes, as their interactions are not naturally simplicial~\cite{mulder2018network, hoppe2025don, Roddenberry2022}. Such models can be formulated in arbitrary dimensions. However, many applications focus on the case where the dynamics is defined on edges and interactions occur via nodes and faces~\cite{Giambagli2022}. Alternative representations of higher-order interactions, such as hypergraphs, have also been explored, but they capture different structural aspects of complex systems~\cite{battiston2020networks,Bick2023}.

The topological Kuramoto model~\cite{millan2020explosive, ghorbanchian2021higher, Nurisso2024UnifiedKuramoto} has become a paradigmatic framework for studying dynamics on complexes, revealing novel synchronization phenomena, including abrupt transitions in higher-order settings~\cite{skardal2020higher, Lucas2020}. Topologically-induced synchronization obstructions have been found in the simplicial Stuart-Landau model, where global topological synchronization has been studied~\cite{carletti2023global}. 
While some aspects of multistability~\cite{Skardal2019multistability,Deville2021} and phase locking~\cite{Nurisso2024UnifiedKuramoto} in similar simplicial models have been analyzed separately, there is still no systematic framework for analyzing how multistable phase-locked states emerge from higher-order interactions.

Here, we develop the \emph{topological nonlinear Kirchhoff conditions} to characterize all phase-locked states in the topological Kuramoto model. These conditions generalize Kirchhoff's current law (KCL) and Kirchhoff's voltage law (KVL) from circuit theory. The generalized KCL yields linear local balance conditions, whose solutions define a continuous family of candidate solutions. The generalized KVL imposes nonlinear global constraints that  ensure that phases are single-valued up to integer multiples of $2\pi$, thereby selecting a finite set of admissible phase-locked states. Based on this approach, we derive a universal boundary-size rule for multistability and uncover that higher-order interactions give rise to structural cascades of coexisting states across dimensions. Using rings, Platonic solids, and simplicial complexes as paradigmatic examples, we demonstrate how topology shapes the existence and organization of phase-locked states, establishing a general framework for synchronization in cell complexes.

\begin{figure*}[tb]
    \centering
    \includegraphics[width=\linewidth]{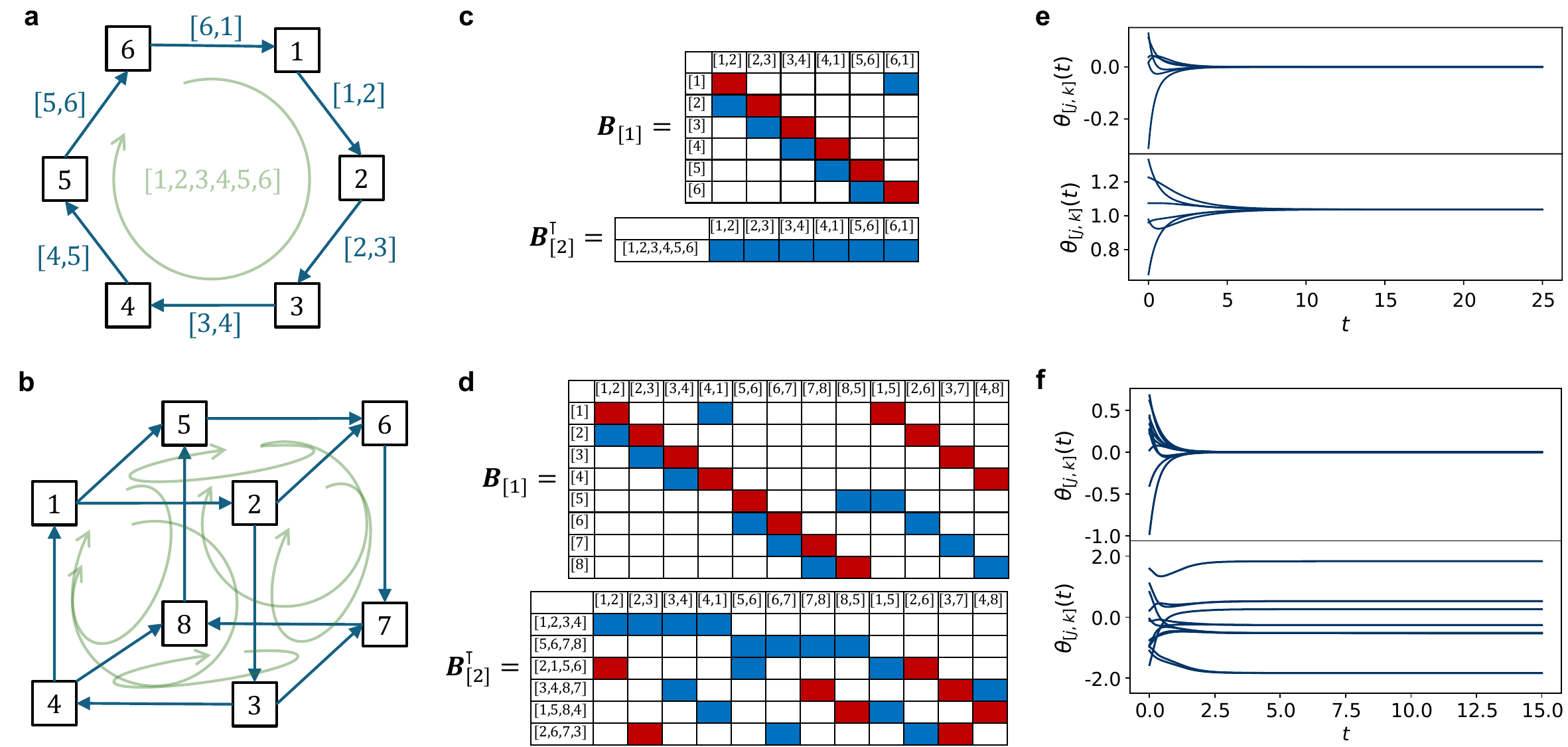}
    \caption{
    \textbf{Examples of cell complexes for $n=1$.} 
    \textbf{a,b} Two representative cell complexes, a ring and a cube, where the dynamics is defined on the edges, and interactions occur via both nodes and faces. Nodes are shown in black, edges in blue and faces in light green. Arrows indicate the orientation used to define the boundary operators.
    \textbf{c,d} Algebraic description of the topology of the cell complexes via boundary operators $\matr{B}_{[1]}$ and $\matr{B}_{[2]}$. 
    Colored fields indicate that a node is in the boundary of an edge, or that an edge is in the boundary of a face. The two colors differentiate a parallel orientation (blue, $+1$) or an anti-parallel orientation (red, $-1$).
    \textbf{e,f} Exemplary time series of the phase variables $\theta_[j,k](t)$ associated to the edges $[j,k]$ for both the ring and the cube. The system converges to different phase-locked states depending on the respective initial conditions.
    }
    \label{fig:examples}
\end{figure*}

\section{Results}

\subsection{Synchronization on graphs and cell complexes}

The Kuramoto model is a paradigmatic model for synchronization in complex networks. Each node $i$ carries a phase variable $\theta_i(t)$, and interactions are mediated by edges $(i,j)$ with coupling strength $K_{ij}$,
\begin{equation}
    \frac{d \theta_i}{dt} = \omega_i - \sum_j K_{ij} \sin(\theta_i - \theta_j).
\end{equation}
In vector notation, the phase variables and natural frequencies are collected in $\vec\theta$ and $\vec\omega$. Phase differences along edges are obtained as $\matr B^\top\vec\theta$, where $\matr B$ is the oriented node-edge incidence matrix, or boundary operator. Its columns correspond to oriented edges and contain two nonzero entries $\pm 1$, marking head and tail. With edge weights collected in a diagonal matrix $\matr K$, the Kuramoto model becomes
\begin{equation}
   \frac{d \vec \theta }{dt} = \vec \omega - \matr B \matr K \sin(\matr B^\top \vec \theta).
\end{equation}
Equivalently, the dynamics of edge phase differences $\vec\theta^{[+]}=\matr B^\top\vec\theta$ is
\begin{equation}
   \frac{d \vec \theta^{[+]} }{dt} = \matr B^\top \vec \omega - \matr B^\top \matr B \, \matr K \sin(\vec \theta^{[+]}),
\end{equation}
where $\matr B^\top\matr B$ is a graph Laplacian.

The standard Kuramoto model is limited to pairwise interactions. The topological Kuramoto model generalizes this setting by assigning phases to cells of arbitrary dimension $n$, such as nodes, edges, faces, or volumes. Interactions between $n$-cell phases occur through adjacent cells of dimension $n-1$ and $n+1$. For $n=0$, the standard Kuramoto model is recovered. For $n=1$, phases are assigned to edges, and two edges interact if they share a node or belong to the same face. Examples of this setting are shown in Fig.~\ref{fig:examples}\textbf{a,b}.

The topology of a cell complex is encoded by boundary operators. The operator $\matr B_{[n]}$ encodes which $(n-1)$-cells lie in the boundary of each $n$-cell. Thus, in the topological Kuramoto model, $\matr B_{[n]}$ describes interactions through lower-dimensional cells, while the coboundary operator $\matr B_{[n+1]}^\top$ describes interactions through higher-dimensional cells. For $n=1$, $\matr B_{[1]}$ is the usual node-edge incidence matrix and $\matr B_{[2]}$ is the edge-face incidence matrix (Fig.~\ref{fig:examples}\textbf{c,d}). The topological Kuramoto model is then written as~\cite{millan2020explosive}
\begin{align}
    \frac{d \vec \theta}{d t} = \vec \omega  
        &-\matr B_{[n]}^\top \matr K_{[n]}  \sin\left( \matr B_{[n]} \vec \theta \right) \nonumber \\
        &- \matr B_{[n+1]} \matr K_{[n+1]} \sin\left( \matr B_{[n+1]}^\top \vec \theta \right).
        \label{eq:eom}
\end{align}
The first term represents the intrinsic frequencies, while the remaining terms capture interactions mediated by adjacent cells of dimension $n-1$ and $n+1$, respectively. We keep the coupling matrices $\matr K_{[n]}$ and $\matr K_{[n+1]}$ general in the theory, but use homogeneous coupling in the numerical examples.

Examples of the dynamics are shown in Fig.~\ref{fig:examples}\textbf{e,f}. For the hexagonal ring and the cube, the phases are associated with edges and interact via nodes and faces. Depending on the initial conditions, the system converges to different stationary phase-locked states. This article aims to characterize all such states.

As in the standard Kuramoto model, it is useful to work not only with the original $n$-cell phases $\vec\theta$, but also with the interaction variables
\begin{align}
\vec \theta^{[-]} = \matr B_{[n]} \vec\theta, 
\qquad
\vec \theta^{[+]} = \matr B_{[n+1]}^\top \vec\theta .
\end{align}
The variable $\vec\theta^{[-]}$ measures phase differences across shared $(n-1)$-cell boundaries, while $\vec\theta^{[+]}$ measures phase accumulation around $(n+1)$-cells. For edge dynamics ($n=1$), these correspond to imbalances at nodes and circulations around faces, respectively.

The equations for these variables decouple because the boundary of a boundary vanishes,
\begin{equation}
    \matr B_{[n]} \matr B_{[n+1]} = 0 
    \qquad \Rightarrow \qquad
    \matr B_{[n+1]}^\top \matr B_{[n]}^\top = 0 .
    \label{eq:boundary-of-boundary}
\end{equation}
For $n=1$, this simply expresses that the oriented boundary of a face has no start or end point: each node appears once with positive and once with negative sign. Multiplying Eq.~\eqref{eq:eom} by $\matr B_{[n+1]}^\top$ and $\matr B_{[n]}$ gives
\begin{align}
    \frac{d \vec \theta^{[+]}}{d t} &=
    \matr B_{[n+1]}^\top \vec \omega
    - \matr L_{[n+1]}^{[\rm down]} \matr K_{[n+1]} 
    \sin\left(  \vec \theta^{[+]}   \right), \\
    \frac{d \vec \theta^{[-]}}{d t} &=
    \matr B_{[n]} \vec \omega
    - \matr L_{[n-1]}^{[\rm up]} \matr K_{[n]} 
    \sin\left(  \vec \theta^{[-]}   \right),
    \label{eq:eom-plus-minus}
\end{align}
with
\begin{align}
    \matr L_{[n-1]}^{[\rm up]}   =
          \matr B_{[n]} \matr B_{[n]}^\top ,
    \qquad
    \matr L_{[n+1]}^{[\rm down]} =
          \matr B_{[n+1]}^\top \matr B_{[n+1]}.
\end{align}
We next recall the basic cell-complex concepts needed to analyze the phase-locked states of these equations.

\subsection{Properties of cell complexes}

Cell complexes are geometrical objects built from elements of different dimension $n$. The building blocks are called $n$-cells and include nodes (0-cells), edges (1-cells), faces (2-cells), volumes (3-cells), and higher-dimensional analogues. Two examples are shown in Fig.~\ref{fig:examples}\textbf{a,b}: a hexagonal face ($n=2$) with six edges and six nodes, and a cube ($n=3$) with six faces, twelve edges, and eight nodes.

An important subclass is given by \emph{simplicial complexes}, where every $n$-cell is an $n$-simplex, i.e., a cell spanned by $n+1$ nodes (e.g., triangles for $n=2$ and tetrahedra for $n=3$). However, many systems involve interactions over loops or surfaces that are not fully connected. Representing such structures as simplicial complexes requires artificial triangulations, introducing additional cells and incidence relations (e.g., a square face must be decomposed into two triangles). We therefore work in the more general setting of cell complexes.

We denote the set of $n$-cells by $S_{[n]}$ and their number by $N_{[n]} = |S_{[n]}|$. For computations, we label cells as $j=1,\dots,N_{[n]}$ and represent them by standard basis vectors $\vec u_j \in \mathbb{R}^{N_{[n]}}$. Vectors defined on $n$-cells are called $n$-cochains.

Boundary operators $\matr B_{[n]} \in \mathbb{R}^{N_{[n-1]} \times N_{[n]}}$ map $n$-cells to their $(n-1)$-dimensional boundaries. After fixing an orientation, their entries take values in $\{-1,0,+1\}$, indicating whether a given $(n-1)$-cell appears in the boundary with positive or negative orientation, or not at all. The choice of orientation determines the signs in $\matr{B}_{[n]}$, but does not imply any physical directionality of the interactions. The transpose $\matr B_{[n]}^\top$ is called the coboundary operator. A fundamental property is that the boundary of a boundary vanishes, expressed by Eq.~\eqref{eq:boundary-of-boundary}. 

The operators $\matr B_{[n]}$ and $\matr B_{[n]}^\top$ can be interpreted as discrete differential operators. In particular, $\matr B_{[n]}$ acts as a discrete divergence, while $\matr B_{[n]}^\top$ acts as a discrete gradient. This interpretation becomes explicit for $n=1$, where $\matr B_{[1]}$ reduces to the node-edge incidence matrix from standard graph theory. Each column of $\matr B_{[1]}$ contains one $+1$ and one $-1$, corresponding to the head and tail of an edge. Thus, $\matr B_{[1]}^\top \vec u$ assigns to each edge the difference of node values, i.e., a discrete gradient. Analogously, $\matr B_{[2]}^\top$ captures circulation around faces and can be viewed as a discrete curl.

The relation $\matr B_{[n]} \matr B_{[n+1]} = 0$ leads to the Hodge decomposition~\cite{barbarossa2020topological}, a discrete analogue of the Helmholtz decomposition from vector calculus. Any vector $\vec x_{[n]} \in \mathbb{R}^{N_{[n]}}$ can be uniquely decomposed as
\begin{align}
\vec{x}_{[n]} = \vec{x}^H_{[n]} + \matr{B}_{[n]}^\top \vec{z}_{[n-1]} + \matr{B}_{[n+1]} \vec{z}_{[n+1]} .
\end{align}
The second term is a gradient-like component induced from lower dimensions, while the third is a curl-like component describing circulation around higher-dimensional cells. In addition, there is a harmonic component $\vec{x}^H_{[n]}$, which satisfies
\[
\matr B_{[n]} \vec{x}^H_{[n]} = \vec{0}, 
\qquad 
\matr B_{[n+1]}^\top \vec{x}^H_{[n]} = \vec{0}.
\]
Thus, it lies in the intersection of the kernels of $\matr{B}_{[n]}$ and $\matr{B}_{[n+1]}^\top$, and its dimension is given by the $n$th Betti number, i.e., the number of independent cycles (holes) in the complex. It corresponds to conserved, divergence-free flows that circulate along topological cycles without being generated by local gradients or curls, and persist due to the presence of holes or loops.

\begin{figure*}[tb]
\centering
\includegraphics[width=\linewidth]{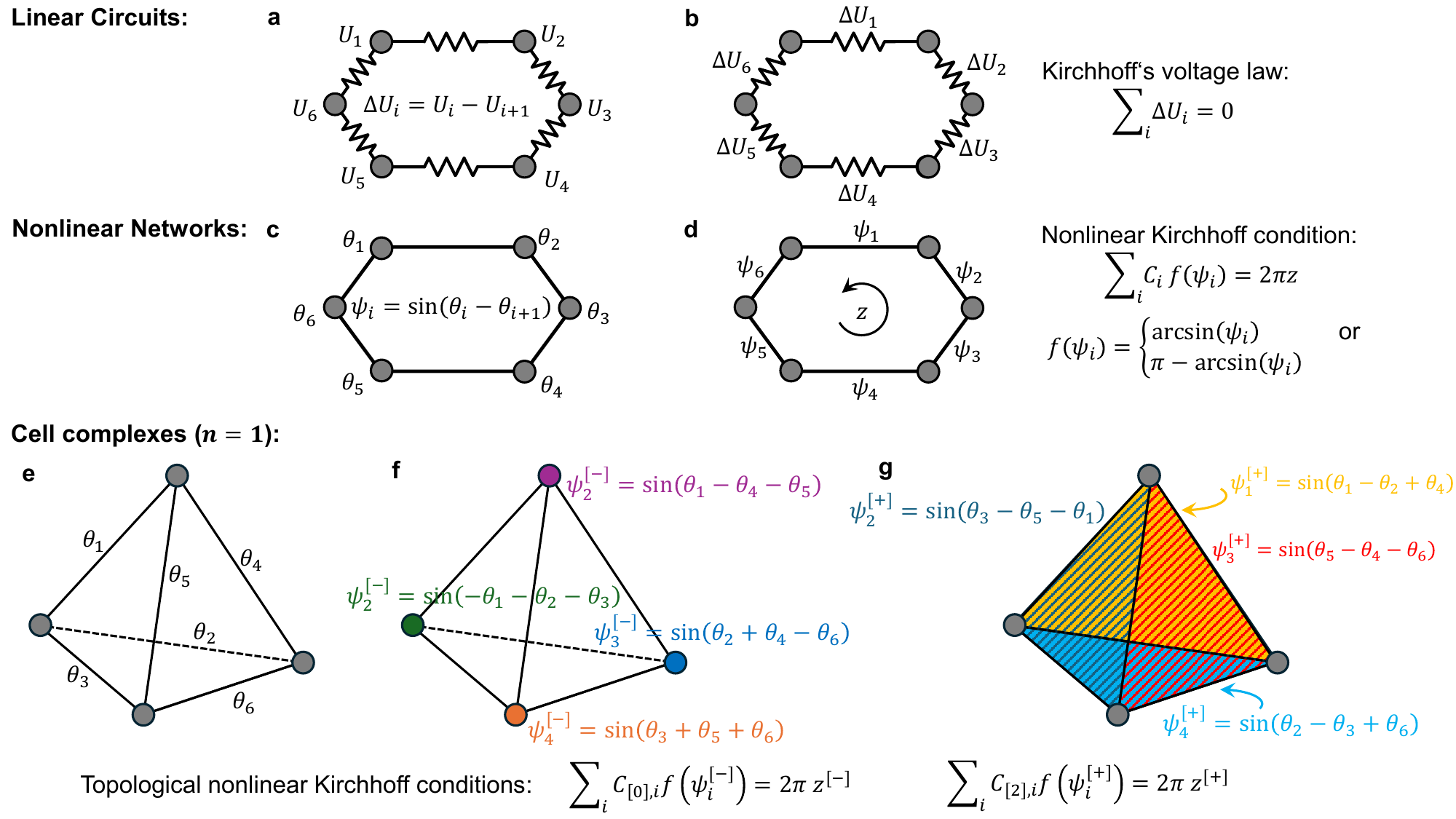}
\caption{
\textbf{From Kirchhoff’s circuit laws to the topological nonlinear Kirchhoff conditions.}  
\textbf{a,b} The analysis of linear electric circuits typically uses nodal voltages $U_j$. However, one can also use currents or voltage differences $\Delta U_i$ across edges which have to obey Kirchhoff’s voltage law (KVL).
\textbf{c,d} The standard Kuramoto model can be described using a nonlinear generalization of the KVL. The fundamental variables are the sines of the phase differences over every edge $\psi_i = \sin(\theta_i-\theta_{i+1})$. In a stationary state, these variables have to satisfy a nonlinear Kirchhoff condition $\sum_i C_i f(\psi_i) = 2\pi z$ with the winding number z.
\textbf{e,f,g} In the topological Kuramoto model, phase variables can be associated to cells of arbitrary dimension $n$ - here we show an example where phases are associated to edges ($n=1$). As fundamental variables, we define the sines of the phase differences $\psi_i^{[\pm]} = \sin(\theta_i^{[\pm]})$ for cells of lower and higher dimension, i.e., for nodes and faces. Both sets of variables must satisfy nonlinear Kirchhoff conditions. 
}
\label{fig:kirchhoff}
\end{figure*}

\subsection{Topological nonlinear Kirchhoff conditions}

We now provide a comprehensive characterization of phase-locked states of the topological Kuramoto model. Our analysis is based on the interaction variables $\vec \theta^{[-]}$ and $\vec \theta^{[+]} = \matr B_{[n+1]}^\top \vec\theta$, whose dynamics is given in Eq.~\eqref{eq:eom-plus-minus}. Setting the time derivatives to zero yields the nonlinear algebraic conditions
\begin{align}
    \matr B_{[n+1]}^\top \vec \omega
    &= \matr L_{[n+1]}^{[\rm down]} \matr K_{[n+1]} 
    \sin\left(  \vec \theta^{[+]}   \right),\label{eq:locked-1} \\
    \matr B_{[n]} \vec \omega
    &=  \matr L_{[n-1]}^{[\rm up]} \matr K_{[n]} 
    \sin\left(  \vec \theta^{[-]}   \right),
    \label{eq:locked}
\end{align}
which define the phase-locked states.

A subtle point is that the conditions $\frac{d}{dt}\vec{\theta}^{[\pm]}=\vec{0}$ do not fully fix the original variables $\vec{\theta}$. In particular, $\vec{\theta}$ may still contain a harmonic component that does not contribute to $\vec{\theta}^{[-]}$ or $\vec{\theta}^{[+]}$. This component affects the dynamics only in a trivial way, corresponding to a global or circulating shift of the phases that does not enter the interaction terms. As a result, it is irrelevant for synchronization, and we refer to solutions of Eqs.~\eqref{eq:locked-1}-\eqref{eq:locked} as phase-locked states rather than strict fixed points.

The linear stability of these states is determined by the Jacobian of Eq.~\eqref{eq:eom} (see Methods). A sufficient condition for stability is
\begin{align}
    \cos\left( \vec \theta^{[-]} \right) > 0,
    \qquad
    \cos\left( \vec \theta^{[+]} \right) > 0.
\end{align}
Under this condition, all non-harmonic perturbations decay, while harmonic perturbations are neutrally stable (see Methods for a proof). Since harmonic modes do not affect the interaction variables $\vec \theta^{[\pm]}$, phase locking is linearly stable. In the following, we focus on such solutions and refer to them as normal phase-locked states.

To characterize all phase-locked states, we adopt a two-step approach. First, we determine candidate solutions by enforcing the local balance conditions in Eqs.~\eqref{eq:locked-1}-\eqref{eq:locked}. These conditions are linear in suitable auxiliary variables and define a low-dimensional solution space, with remaining degrees of freedom corresponding to circulations around cycles. In the second step, we impose nonlinear global consistency conditions that select those solutions that can be realized by actual phase variables.

This approach generalizes Kirchhoff's current and voltage laws, as illustrated in Fig.~\ref{fig:kirchhoff}\textbf{a,b}. A linear electric circuit can be analyzed either in terms of nodal voltages $U_i$ or voltage differences across edges $\Delta U_i$. Kirchhoff's voltage law then requires that the sum of voltage differences along any cycle vanishes.

This framework extends to nonlinear oscillator networks, such as the standard Kuramoto model (Fig.~\ref{fig:kirchhoff}\textbf{c,d}). Phase-locked states can be expressed either in terms of nodal phases $\theta_i$ or auxiliary variables defined as the sines of phase differences across edges, $\psi_i = \sin(\Delta \theta_i)$. The stationarity conditions then become linear in $\psi_i$, while Kirchhoff’s voltage law is replaced by a nonlinear condition involving the inverse sine. Because phases are defined modulo $2\pi$, cycle sums can take values $2\pi z$, where $z\in\mathbb{Z}$ is referred to as the winding number.

For the topological Kuramoto model, this construction must be extended to include interactions via lower- and higher-dimensional cells, as illustrated for an elementary example in Fig.~\ref{fig:kirchhoff}\textbf{e--g}. The auxiliary variables are the sines of the interaction variables $\vec \theta^{[-]}$ and $\vec \theta^{[+]}$. As before, the stationarity conditions become linear, while global consistency imposes nonlinear constraints generalizing Kirchhoff’s voltage law. These \emph{topological nonlinear Kirchhoff conditions} form the central result of this section. In this derivation, we generalize the notion of winding numbers from graph cycles to higher-dimensional cycles of the cell complex.

We now formalize this two-step approach. We introduce auxiliary variables
\begin{align}
    \vec \psi^{[\pm]} = \sin \left( \vec \theta^{[\pm]} \right),
    \label{eq:psi-sine}
\end{align}
which can be interpreted as flow-like variables on the complex, encoding the strength of interactions across cells. In the case $n=1$, the phases $\vec \theta$ are defined on edges, while $\vec \psi^{[+]}$ are associated with faces and $\vec \psi^{[-]}$ with nodes. For $n=0$, the phases are defined on nodes and $\vec \psi^{[+]}$ are defined on edges. In the theory of supply networks, the entries of $\matr K_{[1]} \vec \psi^{[+]}$ are interpreted as flows~\cite{hartmann2024synchronized, Bacic2025phasecohesive, Parameswaran2025symmetrybreaking}.

In the first step, we generalize Kirchhoff's current law and obtain candidate solutions. Using the auxiliary variables, the phase-locking conditions Eq.~\eqref{eq:locked} become
\begin{align}
    \matr B_{[n+1]}^\top \vec \omega
    &= \matr L_{[n+1]}^{[\rm down]} \matr K_{[n+1]} 
    \vec \psi^{[+]}  \label{eq:locked-psi0} \\
    \matr B_{[n]} \vec \omega
    &=  \matr L_{[n-1]}^{[\rm up]} \matr K_{[n]} 
     \vec \psi^{[-]} .
    \label{eq:locked-psi}
\end{align}
Treating $\vec \psi^{[\pm]}$ as independent variables, Eqs.~\eqref{eq:locked-psi0} and~\eqref{eq:locked-psi} form a linear system. The solutions of these equations form a low-dimensional affine subspace.

To parametrize the set of candidate solutions, we compute a basis of the kernels of $\matr B_{[n+1]}$ and $\matr B_{[n]}^\top$. Storing the basis vectors as columns of the matrices $\matr C_{[n+1]}$ and $\matr C_{[n]}^\top$, we have
\begin{align}
    \matr B_{[n+1]} \matr C_{[n+1]} = \matr 0,
    \qquad
    \matr B_{[n]}^\top \matr C_{[n]}^\top = \matr 0.
    \label{eq:kernels}
\end{align}
These matrices are closely related, but not identical, to $\matr B_{[n+2]}$ and $\matr B_{[n-1]}^\top$ via Eq.~\eqref{eq:boundary-of-boundary}. In particular, the columns of $\matr B_{[n+2]}$ ($\matr B_{[n-1]}^\top$) lie in the kernel of $\matr B_{[n+1]}$ ($\matr B_{[n]}^\top$), but the kernel also contains additional harmonic components.

The general solution can therefore be written as
\begin{align}
    \vec \psi^{[+]} &= \vec \psi^{[+]}_{\rm p} 
    +  \matr K_{[n+1]}^{-1} \matr C_{[n+1]} \vec \zeta^{[+]}, 
    \nonumber \\
    \vec \psi^{[-]} &= \vec \psi^{[-]}_{\rm p} 
    +  \matr K_{[n]}^{-1}  \matr C_{[n]}^\top \vec \zeta^{[-]},
    \label{eq:solspace}
\end{align}
where $\vec \zeta^{[\pm]}$ parametrize the space of homogeneous solutions. Mathematically, these variables parametrize contributions from the kernel of the boundary operators, corresponding to harmonic components in the Hodge decomposition. Physically, the parameters $\vec \zeta^{[\pm]}$ quantify the amplitude of circulating flows along independent cycles of the complex. They generalize loop currents in networks: different values of $\vec \zeta^{[\pm]}$ redistribute phase differences around cycles without violating local balance conditions.

The particular solutions $\vec \psi^{[\pm]}_{\rm p}$ can be obtained via the Moore-Penrose pseudoinverse,
\begin{align}
\vec \psi^{[+]}_{\rm p} &= \matr K_{[n+1]}^{-1} \matr B_{[n+1]}^\dagger  \vec \omega, \nonumber \\
\vec \psi^{[-]}_{\rm p} &= \matr K_{[n]}^{-1} \matr B_{[n]}^{\dagger \top} \vec \omega,
\label{eq:sp-sol}
\end{align}
corresponding to the least-squares solution of Eqs.~\eqref{eq:locked-psi0}-\eqref{eq:locked-psi}.

Since $\vec \psi^{[\pm]} = \sin(\vec \theta^{[\pm]})$, the admissible solutions must satisfy
\begin{align}
    | \vec \psi^{[\pm]} | \leq 1.
\end{align}
Thus, the candidate solutions are given by the intersection of an affine subspace with the unit cube, which may be empty depending on $\vec \omega$. In this sense, $\vec\omega$ acts as a control parameter: changing its entries shifts the affine space of candidate solutions and thereby determines which phase-locked states are admissible.

In the second step, we generalize Kirchhoff's voltage law and select valid phase-locked states. In other words, we determine whether a candidate solution \eqref{eq:solspace} can be represented in the form \eqref{eq:psi-sine}. This requires reconstructing $\vec \theta^{[\pm]}$ from $\vec \psi^{[\pm]}$. If Eq.~\eqref{eq:psi-sine} admits a solution, each component has two possible branches, given by $\arcsin(\cdot)$ and $\pi - \arcsin(\cdot)$.

To keep track of these two possibilities, we partition the sets of cells $S_{[n+1]}$ and $S_{[n-1]}$ according to the sign of $\cos \left( \theta^{[\pm]}_i \right)$.  If the cosine is positive, we define the inverse
\begin{align}
    f^{\bullet}(\psi_i^{[\pm]}) = \arcsin(\psi_i^{[\pm]})
\end{align}
up to integer multiples of $2 \pi$ and collect the respective cells in the sets $S_{[n+1]}^{\bullet}$ and $S_{[n-1]}^{\bullet}$. 
Likewise, if the cosine is not positive, we define the inverse
\begin{align}
    f^{\circ}(\psi_i^{[\pm]}) = \pi - \arcsin(\psi_i^{[\pm]})
\end{align}
up to integer multiples of $2 \pi$ and collect the respective cells in the sets $S_{[n+1]}^{\circ}$ and $S_{[n-1]}^{\circ}$.
Finally, we summarize the resulting components in the two vector-valued functions $\vec f_{[n+1]}(\vec \psi^{[+]})$ and $\vec f_{[n-1]}(\vec \psi^{[-]})$. 

The partition is not uniquely determined by the dynamics, but must be chosen consistently with the reconstructed phases. Different partitions correspond to different branches of the inverse sine and therefore to distinct candidate solutions. Physically, the partition specifies how each phase difference is realized, i.e., whether it lies on the increasing or decreasing branch of the sine function, as determined by the sign of $\cos(\theta_i^{[\pm]})$.

Phase-locked states with $S^\circ = \emptyset$ are always linearly stable and are therefore especially relevant for applications. States with $S^\circ \neq \emptyset$ are typically, but not always, unstable. For example, on a ring, choosing all cells in $S^\bullet$ corresponds to states in which all phase differences lie on the branch with $\cos(\theta_i)>0$, i.e., $|\theta_i|<\pi/2$. In contrast, partitions containing elements of $S^\circ$ correspond to configurations where some phase differences lie on the branch with $\cos(\theta_i)\leq 0$. While such states are usually unstable, we will demonstrate an exception where stable phase-locked states with $S^\circ \neq \emptyset$ exist close to bifurcations.

A candidate solution \eqref{eq:solspace} corresponds to a valid phase-locked state if and only if it can be expressed in the form \eqref{eq:psi-sine}. Equivalently, the reconstructed $\vec \theta^{[\pm]}$ must lie in the image of the appropriate boundary or coboundary operators. Since a vector lies in the image of a matrix if and only if it is orthogonal to the kernel of the transpose matrix, this condition can be written as
\begin{align}
   \matr C_{[n+1]}^\top  \vec f_{[n+1]}(\vec \psi^{[+]}) 
   &= 2 \pi \vec z^{[+]},
    \nonumber \\
    \matr C_{[n]}^\top \vec f_{[n-1]}(\vec \psi^{[-]}) 
   &= 2 \pi \vec z^{[-]},
    \label{eq:KVL}
\end{align}
where $\vec z^{[\pm]} \in \mathbb{Z}^m$ are generalized winding vectors. In the case of ordinary graphs ($n=0$), only the winding vectors $\vec z^{[+]}$ exist and its components are referred to as winding numbers~\cite{delabays2016multistability,manik2017cycle,jafarpour2022flow}. 
Equation~\eqref{eq:KVL} enforces global consistency of the reconstructed phase differences. The matrices $\matr C_{[n+1]}^\top$ and $\matr C_{[n]}^\top$ extract sums of phase differences along independent cycles in the corresponding higher- and lower-dimensional cycle spaces. The conditions therefore require that the total phase difference accumulated along each such cycle equals an integer multiple of $2\pi$. The vectors $\vec z^{[\pm]}$ count how many times the phases wind around these cycles, with each component representing the winding number of one independent cycle. Importantly, $\vec z^{[\pm]}=\vec 0$ indicates zero winding, but does not in general imply complete synchronization of the original phase variables $\vec \theta$.

For a fixed partition, Eq.~\eqref{eq:KVL} becomes a nonlinear constraint on the parameters $\vec \zeta^{[\pm]}$ via the parametrization~\eqref{eq:solspace}. It selects those candidate solutions for which there exists a phase vector $\vec \theta$ satisfying $\vec \psi^{[\pm]} = \sin(\vec \theta^{[\pm]})$. In this sense, Eqs.~\eqref{eq:KVL} generalize Kirchhoff’s voltage law: the accumulated phase differences around each independent cycle must be equal to an integer multiple of $2\pi$.

This framework yields the following procedure for computing all phase-locked states.
\begin{alg}{Topological nonlinear Kirchhoff conditions}
\begin{enumerate}
    \item Compute candidate solutions using Eq.~\eqref{eq:solspace}, subject to $|\vec \psi^{[\pm]}| \le 1$.
    \item Choose a partition $S^\bullet, S^\circ$ specifying the branch of the inverse sine for each component.
    \item Solve Eq.~\eqref{eq:KVL} for integer winding vectors $\vec z^{[\pm]}$ consistent with the chosen partition.  
    \item Check linear stability via the Jacobian $\matr J$.
\end{enumerate}\label{eq:alg1}
\end{alg}

Algorithm~\ref{eq:alg1} can be interpreted as a topological and nonlinear generalization of Kirchhoff’s circuit laws. For ordinary graphs ($n=0$), the variables $\matr K_{[1]} \vec \psi^{[+]}$ correspond to currents on edges, and the candidate solutions obtained from Eq.~\eqref{eq:solspace} satisfy Kirchhoff’s current law at each node. The additional constraints in Eq.~\eqref{eq:KVL} enforce a nonlinear version of Kirchhoff’s voltage law, requiring consistency of phase differences around cycles. Compared to classical circuit theory, two key generalizations arise: first, the nonlinearity of the sine leads to multiple solutions associated with different winding numbers; second, the framework extends these Kirchhoff-type conditions to cell complexes of arbitrary dimension.

While related approaches based on Kirchhoff-type formulations and cycle decompositions exist for the standard Kuramoto model on graphs~\cite{delabays2016multistability,manik2017cycle,jafarpour2022flow,Parameswaran2025symmetrybreaking}, these are restricted to pairwise interactions. For the graph case, such methods can be used to derive necessary and sufficient conditions for the existence of phase-locked states in terms of cycle constraints~\cite{hartmann2024synchronized,Bacic2025phasecohesive}. The present framework extends these ideas to higher-order networks, where multiple independent cycles across dimensions contribute to the solution space. This leads to qualitatively new phenomena, including higher-dimensional winding and cascades of multistability, which are absent in pairwise networks.

From a computational perspective, Algorithm~\ref{eq:alg1} provides a constructive procedure for enumerating phase-locked states, building on earlier implementations for graph-based models~\cite{hartmann2024synchronized,Bacic2025phasecohesive}. The code used in this work is available at~\cite{topological_repo2025}.

A key aspect of this procedure is its computational complexity.  In general, the procedure has a combinatorial component due to the choice of partitions $S^\bullet$ and $S^\circ$, whose number scales exponentially as $2^{N_{[n]}}$. This can become prohibitive if one aims to enumerate all phase-locked states.
In many applications, however, one is primarily interested in stable solutions. In this case, it is typically sufficient to restrict to partitions with $S^\circ = \emptyset$, which eliminates the combinatorial complexity associated with enumerating partitions. 
The remaining task is to determine admissible winding vectors $\vec z^{[\pm]}$, set by the topology of the complex, as discussed in the next section.

\subsection{Universal boundary-size rule for multistability}

We now derive a universal criterion for when multistability can arise in the topological Kuramoto model. The topological nonlinear Kirchhoff framework introduced above provides a natural starting point for this analysis, as phase-locked states are organized by winding vectors $\vec z^{[\pm]}$ that encode how phase differences accumulate along independent cycles of the complex.

Phase-locked states correspond to admissible winding vectors $\vec z^{[\pm]}$ that satisfy the nonlinear consistency conditions derived above. The number of phase-locked states is therefore controlled by the admissible values of $\vec z^{[\pm]}$, which are constrained by the topology of the cell complex. In particular, these bounds depend on the number of adjacent cells across dimensions.

More precisely, let $m_k^{[+]}$ denote the number of $(n+1)$-cells in the boundary of the $k$th $(n+2)$-cell, and $m_k^{[-]}$ the number of $n$-cells whose boundary contains the $k$th $(n-1)$-cell. We obtain the universal bounds
\begin{align}
    |z^{[+]}_k| & \le \frac{m_k^{[+]}}{4}, 
     \qquad \mbox{and}  \qquad
    |z^{[-]}_k| \le \frac{m_k^{[-]}}{4},
    \label{eq:windingnumconstraint}
\end{align}
for every $k$ that does not correspond to a harmonic component (see Methods for a proof). In the case of normal phase-locked states with $S^\circ = \emptyset$, the inequalities become strict. This yields a universal boundary-size rule: multistability can only occur if the boundary contains at least five cells, i.e., $m_k^{[\pm]} \ge 5$.

Geometrically, Eq.~\eqref{eq:windingnumconstraint} expresses a phase-budget constraint. Each reconstructed phase difference contributing to a cycle sum is bounded in magnitude by $\pi/2$, while a nonzero winding requires a total phase accumulation of at least $2\pi$. Thus, at least five contributions are required to realize nontrivial winding. Boundaries with fewer than five cells cannot support multistability, whereas larger boundaries can.

\subsection{Rings}
\label{sec:rings}

Rings are among the simplest models of a cell complex, comprised of a single polygonal face bounded by its edges. We start by considering a ring of 6 nodes, 6 edges and 1 face as illustrated in Fig.~\ref{fig:examples}a. Our main interest here will be dynamics of phases defined on the edges (1-cells), which interact via the nodes (0-cells) and face (2-cells). In this example, we demonstrate our algorithm step-by-step, including the study of unstable states.
 
Denoting all cells by the nodes in the respective cell, we have
\begin{align*}
    S_{[0]} &= \big\{ [1],[2],[3],[4] ,[5] ,[6]  \big\} \\
    S_{[1]} &= \big\{ [1,2],[2,3],[3,4],[4,5],[5,6],[6,1]  \big\} \\
    S_{[2]} &= \big\{ [1,2,3,4,5,6] \big\}.
\end{align*}
The boundary operators are given in Fig.~\ref{fig:examples}.

The kernel of the coboundary operator $\matr B_{[1]}^\top$ is spanned by the harmonic vector $\vec 1$ such that $\matr{C}_{[1]}^\top = \vec 1$. The kernel of the boundary operator $\matr B_{[2]}$ is trivial such that we can write $\matr{C}_{[2]} = \vec 0$. Hence, the set of candidate solutions \eqref{eq:solspace} is given by
\begin{align}
    \vec \psi^{[+]} = \vec \psi^{[+]}_{\rm p}, 
    \qquad
    \vec \psi^{[-]} = \vec \psi^{[-]}_{\rm p} 
    +  \matr K_{[1]}^{-1}  \vec 1 \; \zeta^{[-]}.
    \label{eq:solspace-ex1}
\end{align}
For the upper dimension, we always have one $\vec \psi^{[+]} = \vec \psi^{[+]}_{\rm p} = \vec 0$ and we can exclude this dimension from our analysis of multistability. We can thus focus on the lower dimension, for which the condition in Eq.~\eqref{eq:KVL} reads
\begin{align}
    \vec 1^\top \,  \vec f_{[n-1]}(\vec \psi^{[-]}) 
    = 2 \pi \vec z^{[-]}.
    \label{eq:KVL2}
\end{align}

We start by exploring the fully symmetric case when $\vec \omega = 0$ and $\matr K_{[1]} = \eye$. We refer to this case as symmetric because it preserves the full permutation symmetry of the underlying complex. We select the particular solution $\vec \psi^{[-]}_{\rm p} = \vec 0$, such that $\vec \psi^{[-]} =  \zeta^{[-]} \vec 1$ and Eq.~\eqref{eq:KVL2} further simplifies to $\sum_i f_{[n-1],i}(\zeta^{[-]} ) = 2 \pi \vec z^{[-]}$.

At this point, we recall the meaning of the partitions $S_{[0]}^\bullet$ and $S_{[0]}^\circ$: they label the two possible branches of the inverse sine appearing in the stationary equations. On cells in $S_{[0]}^\bullet$ we take $\theta_i^{[-]}=\arcsin(\psi_i^{[-]})$, whereas on cells in $S_{[0]}^\circ$ we take $\theta_i^{[-]}=\pi-\arcsin(\psi_i^{[-]})$. Thus, each partition represents a different candidate branch assignment for reconstructing a phase-locked state from a given solution $\vec\psi^{[-]}$. The physically relevant states will be obtained after solving the nonlinear Kirchhoff conditions.

\begin{figure*}
    \centering
    \includegraphics[width=0.42\paperwidth]{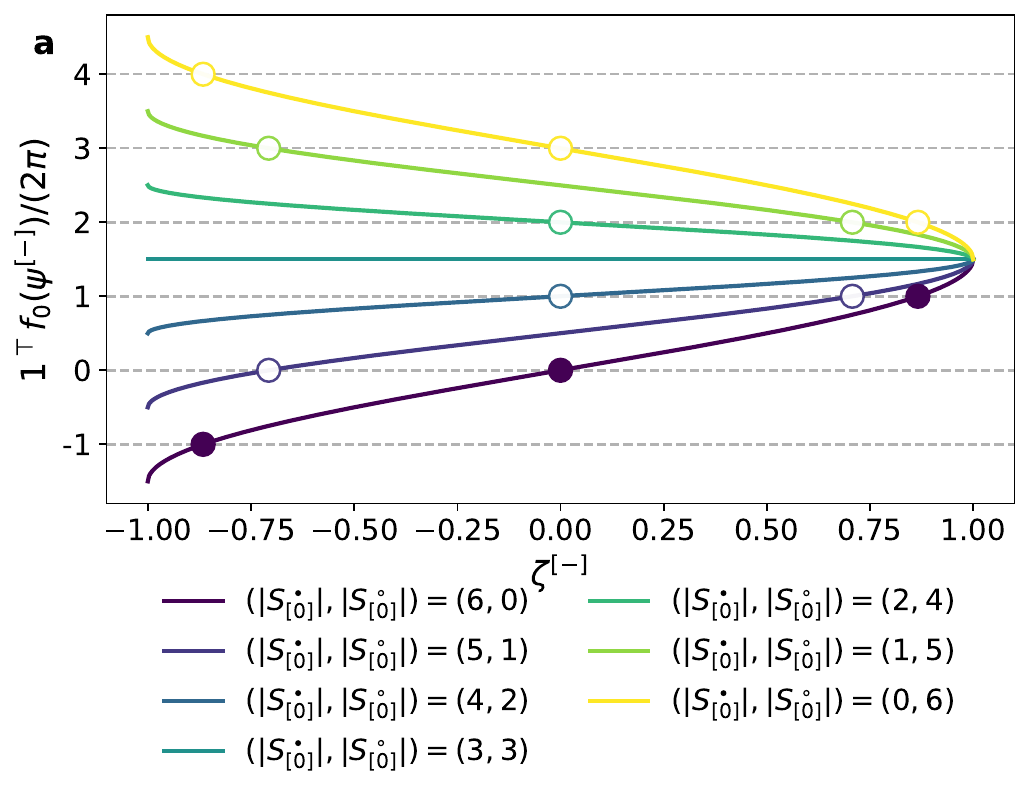}
    \includegraphics[width=0.4\paperwidth]{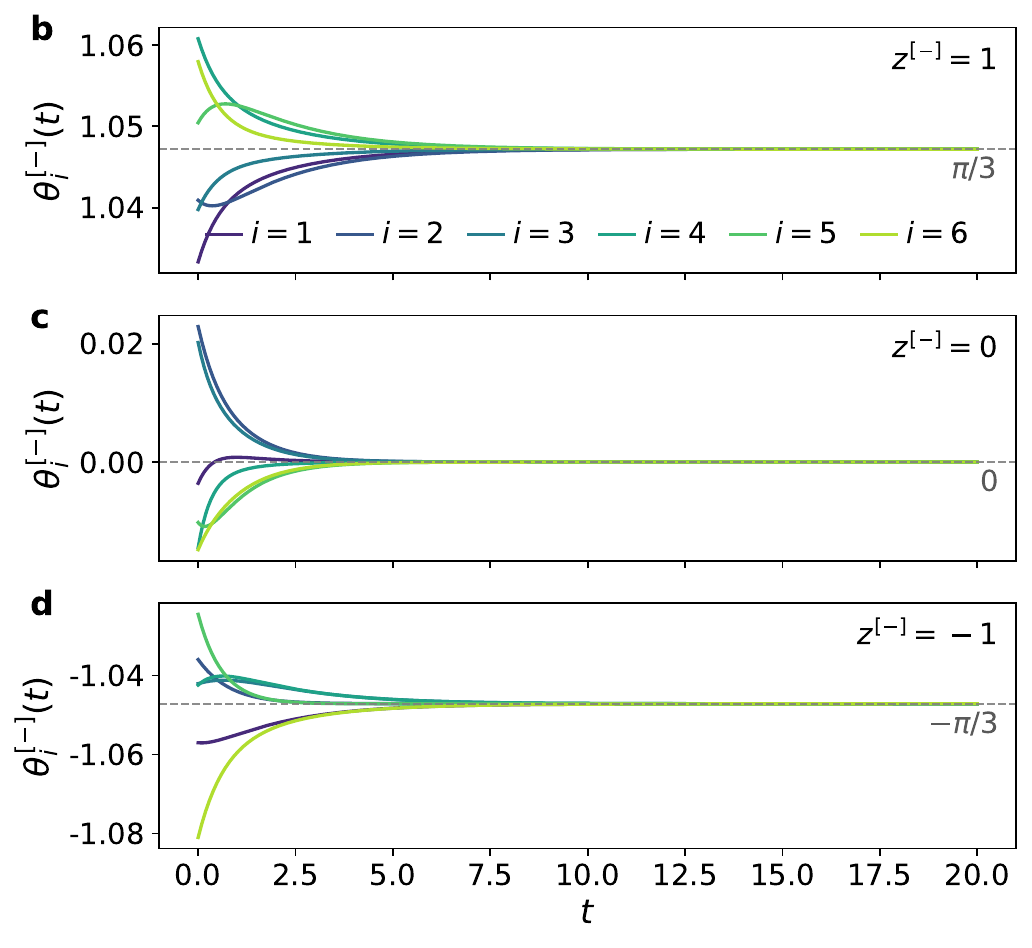}
    \caption{
    \textbf{Phase-locked states of the topological Kuramoto model for a six-edge ring.} \textbf{a} The normalized Kirchhoff condition $C_{[1]}^\top f_{[0]}(\psi^{[-]})/(2\pi)$ as a function of the scalar generalized loop flow amplitude parameter $\zeta^{[-]}$, shown for the distinct multiset partitions of $S_{[0]}=S_{[0]}^\bullet\cup S_{[0]}^\circ$. Intersections with integer values on the vertical axis correspond to phase-locked solutions with winding number $z^{[-]}$. Filled markers denote linearly stable solutions and open markers denote unstable solutions. All partitions except $|S_{[0]}^\bullet|=|S_{[0]}^\circ|=3$ admit phase-locked states. In the all-normal case $|S_{[0]}^\bullet|=6$, $|S_{[0]}^\circ|=0$, three stable solutions coexist, corresponding to winding numbers $z^{[-]}=-1,0,1$. \textbf{b, c, d} Time series $\theta_i^{[-]}(t)$ for the six components of the lower interaction variables, obtained from perturbed initial conditions near the three stable all-normal branches. Dashed horizontal lines indicate the corresponding stationary values $\theta_i^{[-]}=\arcsin(\zeta^{[-]}) = 2\pi z^{[-]}/6$. The trajectories converge to the three coexisting states with winding numbers \textbf{b} $z^{[-]}=-1$, \textbf{c} $z^{[-]}=0$, and \textbf{d} $z^{[-]}=1$.
    }
    \label{fig:example1}
\end{figure*}

In Fig.~\ref{fig:example1} we plot the normalized Kirchhoff condition $\vec 1^\top \,  \vec f_{[n-1]}(\vec \psi^{[-]})/(2\pi)$ as a function of the parameter $\zeta^{[-]}$, for each of the partitions $S_{[0]} = S_{[0]}^\bullet \cup S_{[0]}^\circ$. Since there are six elements that can either be $\bullet$ or $\circ$,
we obtain $2^6=64$ different partitions of $S_{[0]}$. 
Due to symmetry, these 64 curves in  Fig.~\ref{fig:example1} are grouped according to the 7 different multiset permutations (i.e., fixed size of both partitions). Phase-locked states are found where the curves cross an integer value. All of the partitions, except those where $|S_{[0]}^\bullet| = |S_{[0]}^\circ| = 3$ provide phase-locked solutions for $\vec \omega = 0$. 
Finally, after checking the stability of each of the phase-locked solutions, we find multistability for the all-normal partition $|S_{[0]}^\bullet|= 6,  |S_{[0]}^\circ| = 0$, where three stable solutions coexist. All other solutions are unstable.

The three stable solutions correspond to the three twisted states of the hexagonal ring with winding numbers $z^{[-]}=\{-1,0,1\}$. Since all components lie on the normal branch, we have $\theta_i^{[-]}=\arcsin(\psi_i^{[-]})=\arcsin(\zeta^{[-]})$ for every edge $i$. The Kirchhoff condition \eqref{eq:KVL2} then reduces to
\begin{align*}
\sum_{i=1}^6 \theta_i^{[-]} = 2\pi z^{[-]} \implies \theta_i^{[-]}=\frac{2\pi z^{[-]}}{6}, \quad i=1,\dots,6.
\end{align*}
Accordingly, the three stable twisted states with winding numbers $z^{[-]} \in \{-1,0,1\}$ are given by
\begin{align*}
    \zeta^{[-]} &\in \left\{-\sqrt{3}/2,\,0,\,\sqrt{3}/2 \right\}, \\
    \theta_i^{[-]} &\in \left\{-\frac{\pi}{3},\,0,\,\frac{\pi}{3}\right\}.
\end{align*}
In Fig.~\ref{fig:example1}\textbf{b-d} we show time series obtained from perturbed initial conditions near each of the three stable branches, confirming convergence to the twisted states with winding numbers $z^{[-]}=\{-1,0,1\}$.

\begin{figure}
    \centering
    \includegraphics[width=0.98\columnwidth]{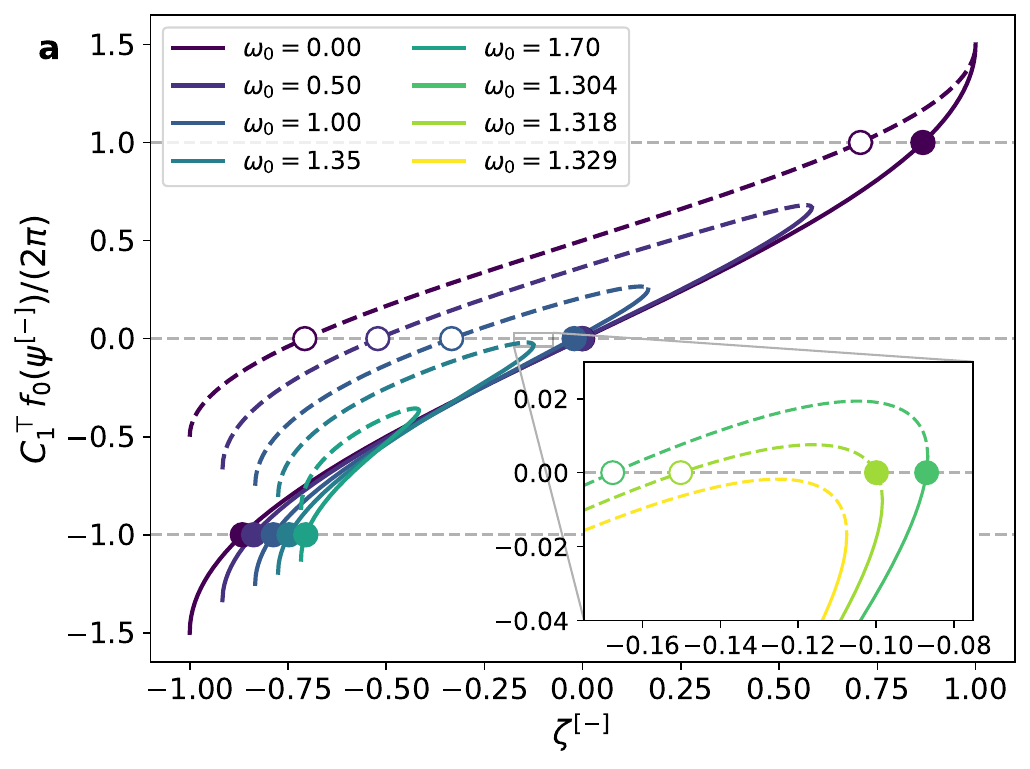}
    \includegraphics[width=\columnwidth]{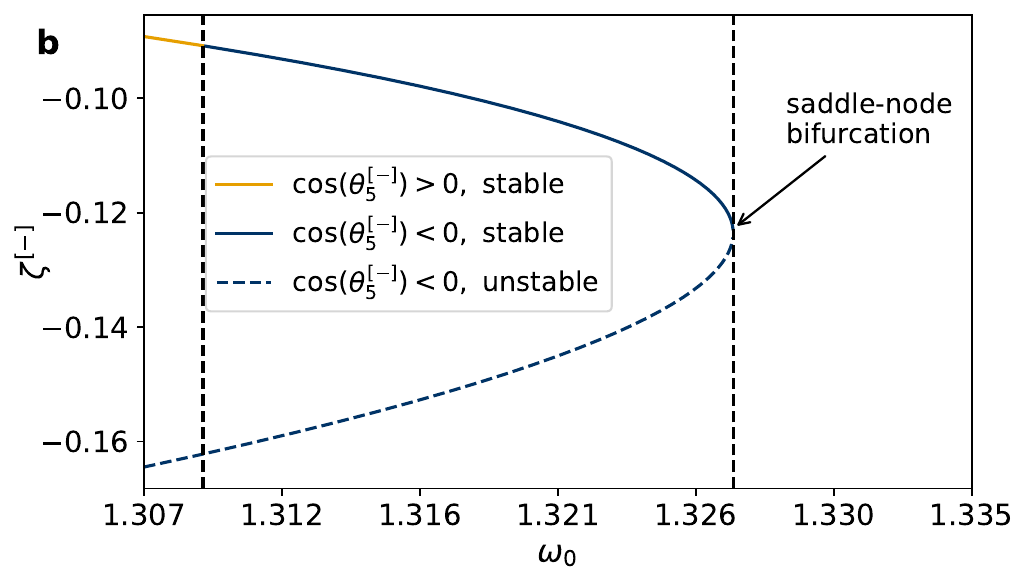}
    \caption{
    \textbf{Effect of heterogeneous intrinsic frequencies on phase-locked states in the six-edge ring.} \textbf{a} Normalized Kirchhoff condition \eqref{eq:KVL2} plotted as a function of the parameter $\zeta^{[-]}$ for two partitions, $S_{[0]}^\circ=\emptyset$, i.e., positive cosine for all lower phase interactions (solid line) and $S_{[0]}^\circ=[5,6]$, i.e., $\cos (\theta_5^{[-]}<0)$ (dashed line), with $\vec\omega=(0,0,0,0,+\omega_0,-\omega_0)$. Colors indicate the value of $\omega_0$. Intersections with integer values on the vertical axis correspond to phase-locked solutions with winding number $z^{[-]}$. Filled and open markers denote linearly stable and unstable solutions. As control parameter $\omega_0$ increases, the number of such intersections decreases, showing the gradual disappearance of phase-locked states. The inset enlarges the neighborhood of $z^{[-]}=0$: in a narrow interval of $\omega_0$, the partition $S_{[0]}^\circ=[5,6]$ admits two distinct solutions, one stable and one unstable, while $S_{[0]}^\circ=\emptyset$ has no intersection. \textbf{b} In this enlarged region, corresponding $z^{[-]}=0$ branches shown as $\zeta^{[-]}$ versus control parameter $\omega_0$. For smaller $\omega_0$, a stable branch with $S_{[0]}^\circ=\emptyset$ and $\cos(\theta^{[-]}_5)>0$ exists. Near $\omega_0\approx1.309$, this branch terminates and a stable branch with $S_{[0]}^\circ=[5,6]$ and $\cos(\theta^{[-]}_5)<0$ appears. At larger $\omega_0$, this stable branch collides with an unstable branch of the same partition in a saddle-node bifurcation near $\omega_0\approx1.327$. The vertical dashed lines mark these two critical values.
    }
    \label{fig:example1asymm}
\end{figure}

Next, we consider heterogeneous natural frequencies, taking 
\begin{align}
   \vec \omega = (0,0,0,0,+\omega_0,-\omega_0).
\end{align}
This choice breaks the underlying permutation symmetry of the ring by singling out one specific cell and we will therefore refer to it as asymmetric. The scalar parameter $\omega_0$ acts as a control parameter for the bifurcation structure shown in Fig.~\ref{fig:example1asymm}.
We limit our exploration to two partitions, $S_{[0]}^\circ = \emptyset$ and $S_{[0]}^\circ = [5,6]$.

In Fig.~\ref{fig:example1asymm}\textbf{a} we plot the Kirchhoff condition \eqref{eq:KVL2}, normalized by $2\pi$, as a function of $\zeta^{[-]}$ for chosen values of $\omega_0$. For small $\omega_0$, the partition $S_{[0]}^\circ = \emptyset$ supports stable phase-locked solutions, whereas the partition $S_{[0]}^\circ = [5,6]$ is typically associated with unstable ones. 

As $\omega_0$ is increased, however, the frequency heterogeneity qualitatively changes the solution structure. In a narrow interval of $\omega_0$, the branch associated with $S_{[0]}^\circ = \emptyset$ disappears, while the partition $S_{[0]}^\circ = [5,6]$ supports two distinct solutions, one stable and one unstable. Thus, heterogeneous intrinsic frequencies act as a control mechanism that selects which partition carries the stable phase-locked state: stability has shifted from the all-normal partition to a mixed partition, $S_{[0]}^\circ=[5,6]$. Increasing $\omega_0$ further, the stable and unstable $S_{[0]}^\circ = [5,6]$ branches collide and annihilate in a saddle-node bifurcation, as shown in Fig.~\ref{fig:example1asymm}\textbf{b}.

Finally, let us generalize our result to polygonal faces of arbitrary size. As before, we assume that the dynamics is defined on the edges (1-cells) and denote the number of edges as $s = N_{[1]}$. We restrict ourselves to the symmetric case $\vec \omega = 0$ and the all-normal partitions $S^\bullet_{[0]}=S_{[0]}$. 

\begin{figure}
    \centering
    \includegraphics[width=0.98\columnwidth]{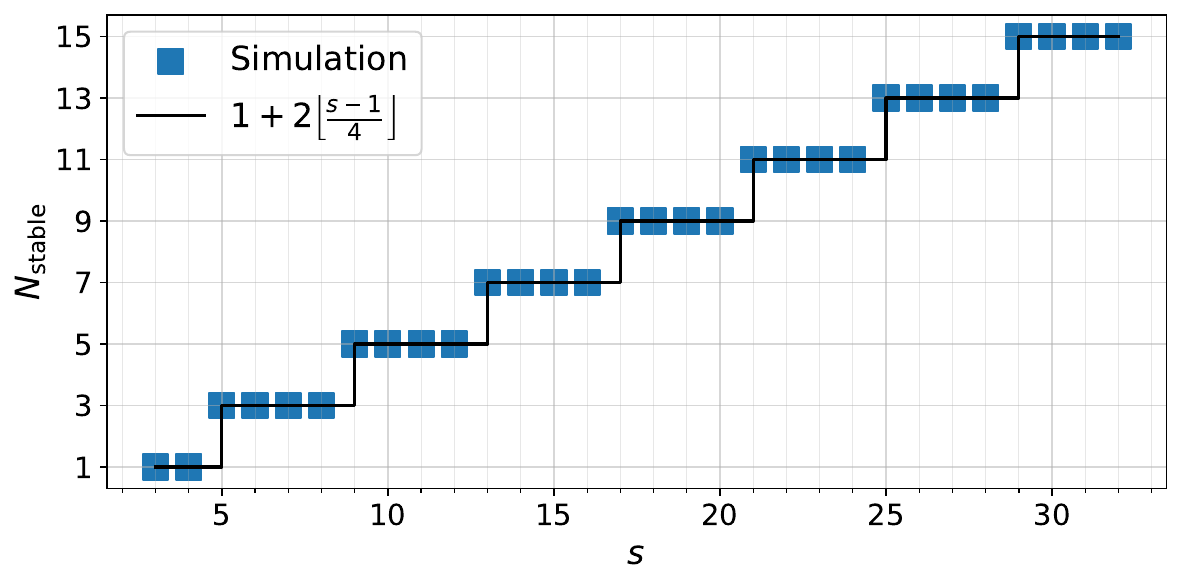}
    \caption{\textit{Dependence of the number of stable phase-locked states on ring size.} Number of stable solutions $N_{\rm stable}$ as a function of ring size $s=N_{[0]}=N_{[1]}$ for the all-normal partition $S_{[0]}^\bullet=S_{[0]}$ and $\vec\omega=\vec 0$. Symbols show the numerically computed number of stable solutions, while the solid line gives the analytical prediction from Eq.~\eqref{eq:z_ring0}.
    }
    \label{fig:Ns_vs_s}
\end{figure}

As in the case of the hexagonal ring, we have $\vec \theta^{[+]} = \vec 0$ and multistability can emerge only through the lower dimension. One finds that twisted states satisfy
\begin{align*}
\theta_i^{[-]}=\frac{2\pi z^{[-]}}{s}, \qquad i=1,\dots,s,
\end{align*}
where $z^{[-]}\in\mathbb Z$ is the winding number. 
For $s \geq 3$, the number of stable phase-locked solutions $N_{\rm stable}$ increases with $s$ as
\begin{align}
    N_{\rm stable} = 1 + 2 \left\lfloor \frac{s - 1}{4} \right\rfloor \label{eq:z_ring0}.
\end{align}
This is a consequence of Eq.~\eqref{eq:windingnumconstraint}. The solution with $z^{[-]} = 0$ always exists, and other solutions appear in pairs as $s$ increases (Fig. \ref{fig:Ns_vs_s}). 
This mirrors the behavior of the standard Kuramoto model~\cite{manik2017cycle} and multistable splay states in ring-like oscillator systems~\cite{gurevich2025multistable}.

\begin{figure*}
    \centering
    \includegraphics[width=0.83\paperwidth]{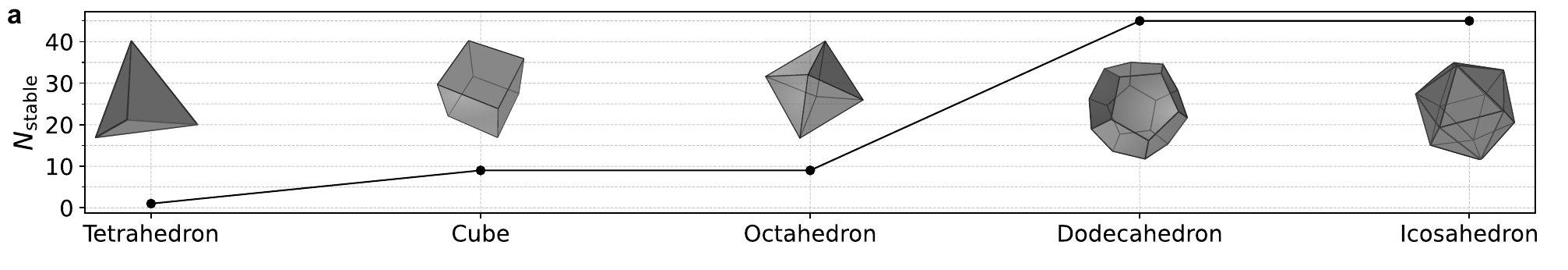}
    \includegraphics[width=0.83\paperwidth]{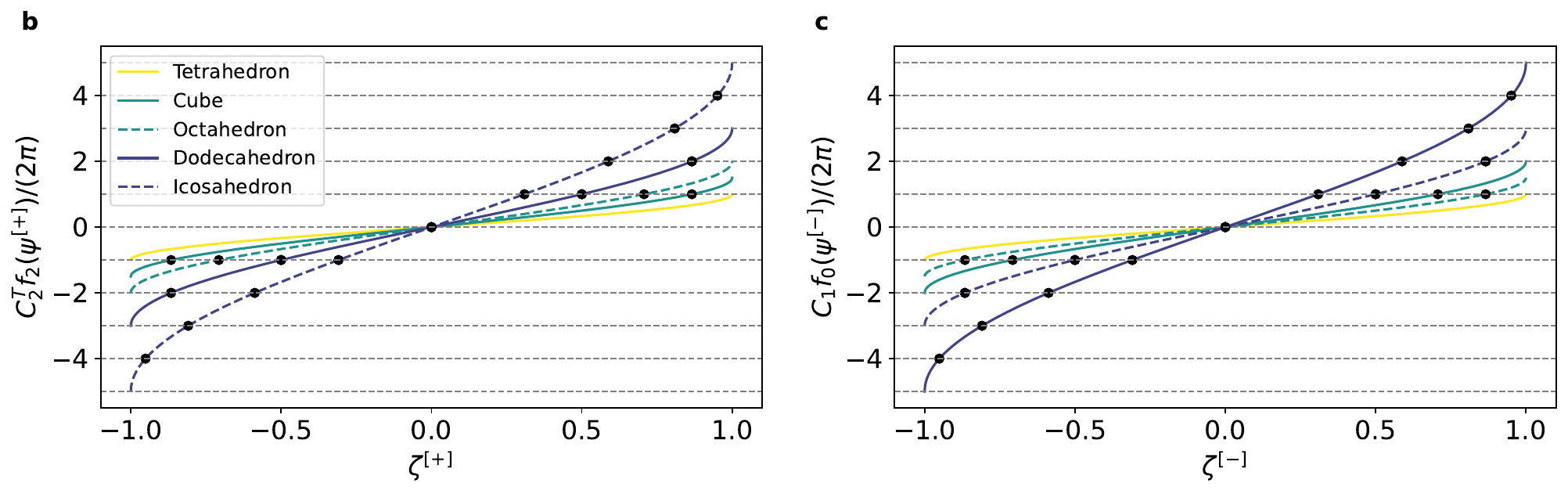}
    \includegraphics[width=0.83\paperwidth]{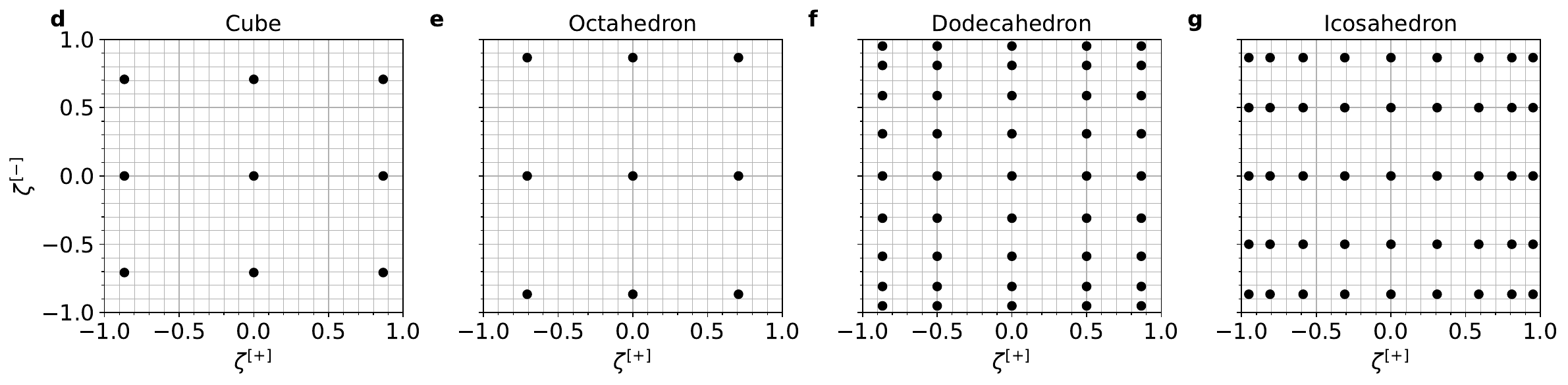}
    \caption{\textbf{Multistability in Platonic solids (edge dynamics, $n=1$).}
    \textbf{a} Number of stable all-normal phase-locked states $N_{\mathrm{stable}}$ for the five Platonic solids at $\vec{\omega}=\vec{0}$. \textbf{b, c} Normalized Kirchhoff sums $C_{[2]}^{\top}f_{[2]}(\psi^{[+]})/(2\pi)$ and $C_{[1]}f_{[0]}(\psi^{[-]})/(2\pi)$, plotted as functions of $\zeta^{[+]}$ and $\zeta^{[-]}$, respectively. Intersections with integer values correspond to phase-locked solutions of Eq.~\eqref{eq:KVL}; filled markers denote stable solutions. Except for the tetrahedron, all Platonic solids exhibit cascades of multistability associated with non-trivial winding in both lower- and higher-dimensional sectors. Dual solids are shown in matching colors and distinguished by line style. \textbf{d, e, f, g} Stable phase-locked solutions in the $(\zeta^{[+]},\zeta^{[-]})$ plane for the cube, octahedron, dodecahedron, and icosahedron.}
    \label{fig:platonic_solids}
\end{figure*}

\begin{figure*}
    \centering
    \includegraphics[width=0.83\paperwidth]{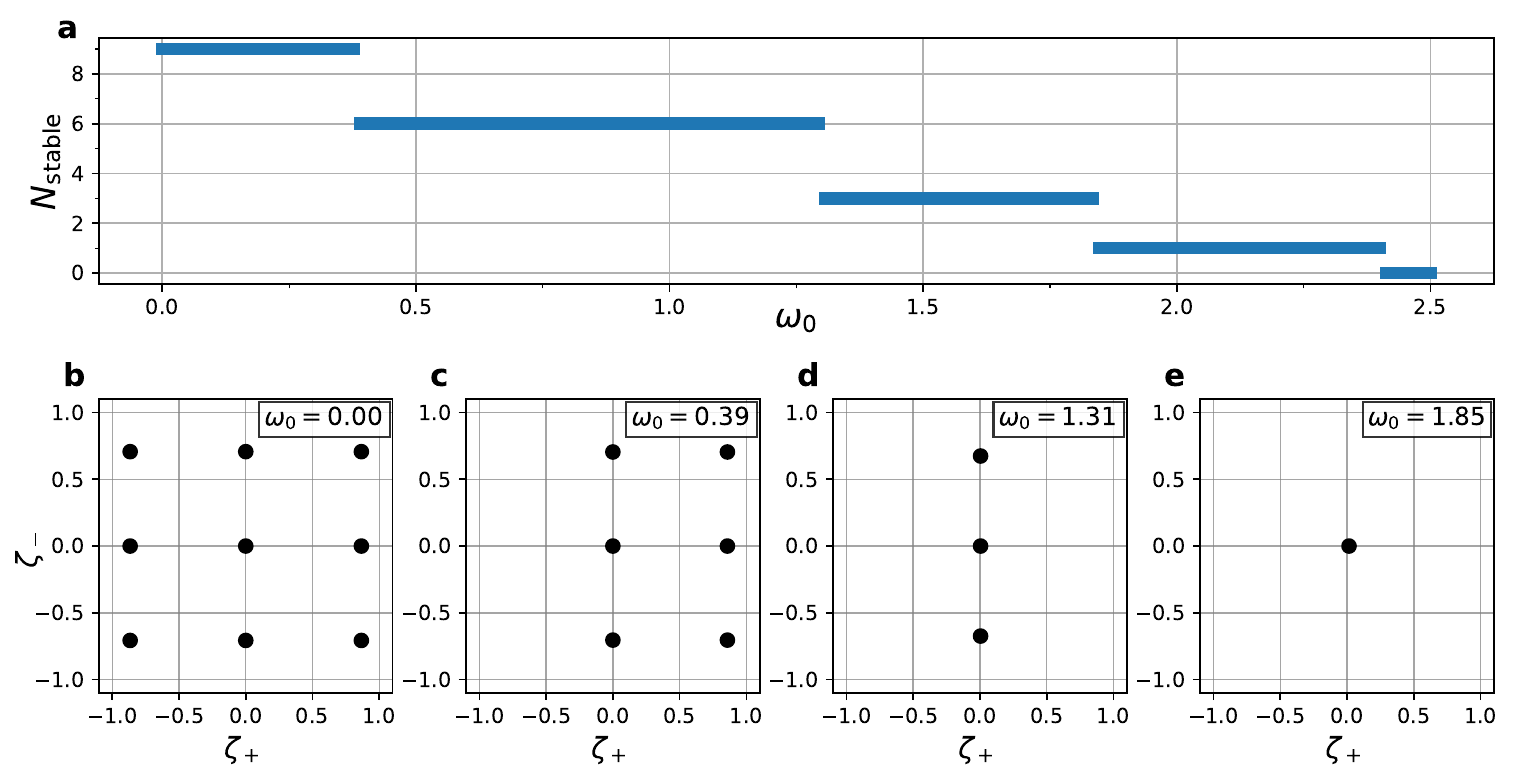}
    \caption{
    \textbf{Control of the multistability cascade in the cube by heterogeneous intrinsic frequencies.}
    \textbf{a} Number of stable phase-locked states \(N_{\mathrm{stable}}\) for the all-normal partition of the cube as a function of the control parameter $\omega_0$, with $\vec \omega = (0,0,0,0,0,0,0,0,+\omega_0,0,0,+\omega_0)$. Increasing $\omega_0$ progressively removes stable branches and thereby reverses the multistability cascade present in the symmetric case $\vec\omega=\vec 0$. \textbf{b, c, d, e} Corresponding stable and unstable phase-locked solutions in the $(\zeta^{[+]},\zeta^{[-]})$ plane for selected values of $\omega_0$. Filled markers denote stable solutions. Together, the panels show how $\omega_0$ acts as a control parameter for the number of stable phase-locked states.}
    \label{fig:example2-fp}
\end{figure*}

\subsection{Platonic solids}
\label{sec:solids}

We now consider convex regular polyhedra in three-dimensional space, known as the Platonic solids: tetrahedron, cube, octahedron, dodecahedron, and icosahedron (Fig.~\ref{fig:platonic_solids}). Each is treated as a cell complex with nodes, edges, faces, and volume. While the tetrahedron, octahedron, and icosahedron are simplicial complexes (all faces triangular), the cube and dodecahedron are not.

We focus on edge dynamics ($n=1$), where interactions occur via nodes and faces. The number of possible partitions grows exponentially, so we restrict attention to the all-normal partition $S_{[2]}^\bullet = S_{[2]}$ and $S_{[0]}^\bullet = S_{[0]}$, which yields stable solutions, and consider the homogeneous case $\vec \omega = \vec 0$.

In all five solids, the relevant kernels are one-dimensional and spanned by $\vec 1$. Hence, the candidate solutions simplify to
\begin{align*}
  \vec \psi^{[+]} = \zeta^{[+]} \vec 1, \qquad
  \vec \psi^{[-]} = \zeta^{[-]} \vec 1,
\end{align*}
and, for the all-normal partition,
\begin{align*}
  \vec \theta^{[+]} = \arcsin(\zeta^{[+]}) \vec 1, \qquad
  \vec \theta^{[-]} = \arcsin(\zeta^{[-]}) \vec 1,
\end{align*}
and the nonlinear Kirchhoff conditions reduce to
\begin{align*}
    N_{[2]} \arcsin(\zeta^{[+]}) = 2 \pi z^{[+]}, \qquad
    N_{[0]} \arcsin(\zeta^{[-]}) = 2 \pi z^{[-]},
\end{align*}
yielding
\begin{align*}
    \vec \theta^{[+]} &= \frac{2 \pi z^{[+]}}{N_{[2]}} \vec 1, \qquad
    \vec \theta^{[-]} = \frac{2 \pi z^{[-]}}{N_{[0]}} \vec 1.
\end{align*}

Thus, normal phase-locked states are parameterized by the Cartesian product of the admissible winding numbers 
$z^{[+]}$ and $z^{[-]}$, associated with faces and nodes, respectively, and their total number is 
\begin{align}
\label{Nst_tot}
N_{\mathrm{stable}} =
\left( 1 + 2 \left\lfloor \frac{N_{[0]} - 1}{4} \right\rfloor \right)
\left( 1 + 2 \left\lfloor \frac{N_{[2]} - 1}{4} \right\rfloor \right).
\end{align}

For the tetrahedron ($N_{[0]}=N_{[2]}=4$), only the trivial state exists. For larger solids, additional winding numbers become admissible, leading to multistability. For example, the cube and octahedron yield $N_{\mathrm{stable}}=9$, while the dodecahedron and icosahedron yield $N_{\mathrm{stable}}=45$.

Equation~\eqref{Nst_tot} shows that the number of stable states depends only on $(N_{[0]},N_{[2]})$. This explains why dual solids (cube/octahedron and dodecahedron/icosahedron) exhibit identical multistability: duality exchanges $N_{[0]}$ and $N_{[2]}$, leaving $N_{\mathrm{stable}}$ invariant.

Figure~\ref{fig:platonic_solids} illustrates how admissible winding numbers are selected independently in the lower- and higher-dimensional sectors, leading to grid-like patterns of solutions. Except for the tetrahedron, all Platonic solids exhibit cascades of multistability arising from both dimensions.

We finally demonstrate how heterogeneity controls these states using the cube. For $\vec \omega = \vec 0$, nine stable states coexist. Introducing heterogeneous frequencies breaks symmetry and progressively eliminates solutions, with states disappearing in symmetric pairs or triplets (Fig.~\ref{fig:example2-fp}). Thus, intrinsic heterogeneity controls the number of stable phase-locked states.

\subsection{Simplicial complexes generated by a single simplex}

We now consider simplicial complexes generated by a single $d$-simplex, i.e., a $d$-simplex together with all of its boundary simplices of dimensions $n=d-1,d-2,\ldots,0$. For example, a tetrahedron (3-cell) gives rise to a simplicial complex consisting of its faces (2-cells), edges (1-cells), and nodes (0-cells). We then fix a dimension $n \le d-1$ and analyze phase-locked states of the topological Kuramoto model defined on $n$-simplices interacting via simplices of dimension $n \pm 1$. 

In low dimensions, such complexes admit only limited multistability. The winding number constraint in Eq.~\eqref{eq:windingnumconstraint} implies that nontrivial multistability requires boundaries with more than four cells. Accordingly, the 2-simplex (triangle) and 3-simplex (tetrahedron) each admit only a single stable solution for $n=1$.

We focus on the stable all-normal solutions, $S_{[n+1]}^\bullet = S_{[n+1]}$ and $S_{[n-1]}^\bullet = S_{[n-1]}$, for simplicial complexes generated by a single $d$-simplex, for all admissible $n$ up to $d=5$. We set $\vec \omega=0$ and all couplings to unity. The explicit boundary operators and kernel matrices are provided in~\cite{topological_repo2025}. Results are summarized in Table~\ref{tab:Nst_simplexes_full}, 
listing the cell counts, the admissible cochain dimension $n$, the number of stable all-normal states, and the corresponding winding vectors. The winding vectors characterize topological winding in the interaction variables $\vec\theta^{[\pm]}$. Each block of rows corresponds to one simplicial complex, while the individual rows within a block distinguish the different values of $n$. Whenever $N_{\mathrm{stable}}=1$, the unique stable state has trivial winding numbers. The table shows that nontrivial multistability first appears only for the 4-simplex and 5-simplex, and only for selected values of $n$.

\begin{table*}[t]
\centering
\begin{tabular*}{\textwidth}{@{\extracolsep{\fill}} c c c c}
\hline
Complex $(N_{[0]},N_{[1]},\dots)$ & $n$ & $N_{\mathrm{stable}}$ & Stable winding numbers \\
\hline
\multirow{2}{*}{\shortstack{2-simplex (triangle): $(3,3)$}}
& 0 & 1 & $z^{[+]} = 0$ \\
& 1 & 1 & $z^{[-]} = 0$ \\
\hline
\multirow{3}{*}{\shortstack{3-simplex (tetrahedron): $(4,6,4)$}}
& 0 & 1 & $\vec z^{[+]} = \vec 0_{3}$ \\
& 1 & 1 & $z^{[-]} = 0$, \ $z^{[+]} = 0$ \\
& 2 & 1 & $\vec z^{[-]} = \vec 0_{3}$ \\
\hline
\multirow{4}{*}{\shortstack{4-simplex (5-cell): $(5,10,10,5)$}}
& 0 & 1 & $\vec z^{[+]} = \vec 0_{6}$ \\
& 1 & \textbf{3} & $\vec z^{[+]} = \vec 0_{4}$, \ $z^{[-]} \in \{0,\pm1\}$ \\
& 2 & \textbf{3} & $z^{[+]} \in \{0,\pm1\}$, \ $\vec z^{[-]} = \vec 0_{4}$ \\
& 3 & 1 & $\vec z^{[-]} = \vec 0_{6}$ \\
\hline
\multirow{5}{*}{\shortstack{5-simplex: $(6,15,20,15,6)$}}
& 0 & 1 & $\vec z^{[+]} = \vec 0_{10}$ \\
& 1 & \textbf{3} & $\vec z^{[+]} = \vec 0_{10}$, \ $z^{[-]} \in \{0,\pm1\}$ \\
& 2 & 1 & $\vec z^{[+]} = \vec 0_{5}$, \ $\vec z^{[-]} = \vec 0_{5}$ \\
& 3 & \textbf{3} & $z^{[+]} \in \{0,\pm1\}$, \ $\vec z^{[-]} = \vec 0_{10}$ \\
& 4 & 1 & $\vec z^{[-]} = \vec 0_{10}$ \\
\hline
\end{tabular*}
\caption{\textbf{Stable phase-locked states for simplicial complexes generated by a single $d$-simplex for $d\leq 5$.} The first column lists each simplicial complex together with the number of cells of different dimension. For each admissible cochain dimension $n$,
the table gives the number of stable all-normal states and the associated winding numbers. Here $\vec 0_m$ denotes the zero vector of dimension $m$. Entries with more than one stable state are shown in bold.}
\label{tab:Nst_simplexes_full}
\end{table*}

For arbitrary $d$, the dimension of $\vec z^{[\pm]}$ is
\begin{align}
  \dim \vec z^{[+]} &= \dim \ker(\matr B_{n+1}) = \binom{d}{n+2} 
  \nonumber
  \\
  \dim \vec z^{[-]} &= \dim \ker(\matr B_n^\top) = \binom{d}{n-1}
  \label{eq:dim-d-n}
\end{align}
with $\dim \vec z^{[-]} = 0$ for $n=0$ (see Methods for a proof). The elements of $\vec z^{[\pm]}$ are integers. One can then apply equations \eqref{eq:windingnumconstraint} to compute upper bounds for the number of allowed winding numbers. Deciding which vectors $\vec z^{[\pm]}$ are achievable is assisted by symmetry arguments: simplices with $\vec \omega = \vec 0$ and equal coupling are symmetric to node permutations. Hence, if a certain $\vec z^{[\pm]}$ is achievable, then any permutation of its components, should also be achievable.

\section{Discussion}

Our results establish a general theoretical framework for studying multistability in oscillator networks with higher-order interactions, integrating concepts from network science, geometry, and nonlinear dynamics. The topological nonlinear Kirchhoff conditions algorithm allows one to identify all phase-locked states of the topological Kuramoto model on arbitrary cell complexes. By applying it to minimal motifs (polygonal rings, Platonic solids, and simplexes), we find that multistability emerges when boundaries contain more than four cells. Structural cascades of multistability then arise, with stable states inherited from lower and/or upper dimensions. Moreover, dual Platonic solids display identical multistability patterns, suggesting the existence of universality classes determined by the boundary structure. By clarifying how multistability emerges from the topology and boundary structure of interactions, our framework provides a principled tool to predict and control synchronization patterns and potential failure modes in complex systems, including oscillator networks in neuroscience and engineered infrastructures such as power grids, where cycle structure and flow constraints play a central role.

Future research could explore how these structural cascades manifest in heterogeneous systems, and how combining the elementary motifs studied here gives rise to multistability in larger complexes. Beyond theoretical interest, these findings are relevant for real-world systems where higher-order couplings are intrinsic, such as neuronal networks ~\cite{Breakspear2010}, optical and mechanical oscillator arrays ~\cite{Nixon2013, Zhang2012}, power-grid stability ~\cite{dorfler2014synchronization, Motter2013}, and collective decision-making in multi-agent systems~\cite{OlfatiSaber2007}. The framework may be applicable to other dynamical processes on cell complexes, including diffusion, consensus, and pattern-forming instabilities.

\section{Methods}

\subsection{Stability of phase-locked states}
\label{app:eigen}

The stability of phase locking can be determined from the Jacobian matrix
of the equations of motion \eqref{eq:eom} given by 
\begin{align}
    \matr J
    = & -\matr B_{[n]}^\top \matr K_{[n]}  
     \mbox{diag} \left[ \cos\left( \vec \theta^{[-]} \right) \right]  \matr B_{[n]} 
     \nonumber \\
     &- \matr B_{[n+1]} \matr K_{[n+1]} 
     \mbox{diag}\left[ \cos\left( \vec \theta^{[+]} \right) \right] \matr B_{[n+1]}^\top.
     \label{eq:jacobian}
\end{align}

In general, a stationary state is linearly stable if the real part of all eigenvalues is negative~\cite{strogatz2018nonlinear}. If the stationary state satisfies the conditions
\begin{align}
    \cos\left( \vec \theta^{[-]} \right)
    > 0 
    \quad \mbox{and} \quad
    \cos\left( \vec \theta^{[+]}\right) 
    > 0
    \label{eq:normal2}
\end{align}
we find that all eigenvalues are real and satisfy the following conditions: (i) If an eigenvector $\vec v$ is purely harmonic, then the associated eigenvalue is given by $\lambda = 0$. (ii) If an eigenvector $\vec v$ is not purely harmonic, then the associated eigenvalue satisfies $\lambda < 0$, as proven below . We thus conclude that a stationary state that satisfies Eq.~\eqref{eq:normal2} is linearly stable with respect to perturbations that are orthogonal to the harmonic subspace and neutrally stable with respect to harmonic perturbations. Since harmonic perturbations do not affect the projected quantities $\vec \theta^{[+]}$ and $\vec \theta^{[-]}$, phase locking is linearly stable.

Proof:
Let $\vec v \neq \vec 0$ be an eigenvector of the Jacobian matrix $\matr J$ with associated eigenvalue $\lambda$ such that
\begin{align*}
    &\matr J \vec v = \lambda \vec v
    \qquad \Rightarrow \qquad
    \lambda = \frac{\vec v^\top \matr J \vec v}{\vec v^\top \vec v} \, .
\end{align*}
If $\vec v$ is purely harmonic then $\matr B_{[n]} \vec v = \vec 0$ and $\matr B_{[n+1]}^\top \vec v = 0$ by definition and thus $\matr J \vec v = \vec 0 \Rightarrow \lambda = 0.$
If $\vec v$ is not purely harmonic then
$\vec y^{[-]}  = \matr B_{[n]} \vec v \neq 0$ or/and $\vec y^{[+]} = \matr B_{[n+1]}^\top \vec v \neq 0$. We thus obtain
\begin{align*}
    \lambda = & - (\vec v^\top \vec v )^{-1} \bigg\{
        \vec y^{[-]\top} \matr K_{[n]} 
     \mbox{diag}\left[ \cos\left( \vec {\bar \theta}^{[-]} \right) \right]    \vec y^{[-]} 
     \\
     & \qquad  \qquad \qquad
     +  \vec y^{[+]\top } \matr K_{[n+1]} 
     \mbox{diag}\left[ \cos\left( \vec {\bar \theta}^{[+]} \right) \right]    \vec y^{[+]} 
    \bigg\}
    \\
    &= - (\vec v^\top \vec v )^{-1} \bigg\{
        \sum_j \left( y_j^{[-]} \right)^2 (K_{[n]})_{jj}  \cos\left(  {\bar \theta}^{[-]}_j \right) 
    \\
    & \qquad \qquad \qquad
       + \sum_j \left( y_j^{[+]} \right)^2 (K_{[n+1]})_{jj}  \cos\left(  {\bar \theta}^{[+]}_j \right) 
    \bigg\}.
\end{align*}
If the stationary state satisfies the conditions in Eq.~\eqref{eq:normal2}, this implies $\lambda < 0$.

\subsection{Proof of Eq.~\eqref{eq:windingnumconstraint}}

To prove Eq.~\eqref{eq:windingnumconstraint}, we rewrite Eq.~\eqref{eq:KVL} in components and use that  $|f_\pm| \le \pi/2$ and obtain
\begin{align*}
    |z^{[+]}_k| & = \frac{1}{2\pi}
    \left| \sum_{i}  C_{[n+1],ik} f_{\pm}(\psi_i^{[+]}) \right|
    \le \frac{1}{4} \sum_{i} |C_{[n+1],ik}| 
    \\
    |z^{[-]}_k| &= \frac{1}{2\pi}
    \left| \sum_{i} C_{[n],ki} f_{\pm}(\psi_i^{[-]}) \right|
    \le \frac{1}{4} \sum_{i} |C_{[n],ki} | .
\end{align*}
In the case of normal phase-locked states with $S^\circ = \emptyset$ we obtain a $<$ instead of the $\le$.

We recall that the matrices $\matr C$ are closely related to the boundary operators. We can choose a basis of the kernel such that the $k$th column of $\matr C_{[n+1]}$ ($\matr C_{[n]}^\top$) is equal to the $k$th column of $\matr B_{[n+2]}$ ($\matr B_{[n-1]}^\top$) unless $k$ corresponds to the harmonic component. 
A further simplification is possible by noting that the entries of the boundary operator have the values $-1,0,+1$. We then define $m_k^{[+]}$ as the number of non-zero elements of the $k$th column of $\matr B_{[n+2]}$ and $m_k^{[-]}$ as the number of non-zero elements of the $k$th row of $\matr B_{[n-1]}$. We then obtain the bounds given in Eq.~\eqref{eq:windingnumconstraint} for every $k$ that does not correspond to the harmonic component.

Notably, the number of non-zero elements can be related to the number of adjacent cells and thus enables a geometric interpretation. The number $m_k^{[+]}$ counts the number of cells of dimension $n+1$ in the boundary of the $k$th cell of dimension $n+2$. The winding number $z_k^{[+]}$ counts how often the phases $\vec \theta^{[+]}$ wind by $2\pi$ when following all cells in this boundary.  Likewise, the number $m_k^{[-]}$ counts the number of cells of dimension $n$ that have the $k$th cell of dimension $n-1$ in its boundary.

\subsection{Proof of equations~\eqref{eq:dim-d-n}}

We prove the equations \eqref{eq:dim-d-n} using a recurrence relation that is derived from the fact that $d$-simplex is contractible and the rank-nullity theorem. A $d$-simplex is contractible, which means that all homology groups are trivial except $H_0 \cong \mathbb{R}$,
\begin{align}
    H_n = 0, \quad n>0,
\end{align}
where a homology group is the quotient vector space
\begin{align}
H_n = \frac{\ker \matr B_{[n]}}{\text{im } \matr B_{[n+1]}}
\end{align}
implying
\begin{align}
\dim H_n = \dim \ker \matr B_{[n]} - \dim \text{im } \matr B_{[n+1]}.
\end{align}
Since $\matr B_{[n]} \matr B_{[n+1]} = 0$, every element in $\matr B_{[n+1]}$ lies in $\ker \matr B_{[n]}$,
\begin{align}
\text{im } \matr B_{[n+1]} \subseteq \ker \matr B_{[n]} \label{eq:subset}
\end{align}
For $n>0$, since $H_n = 0$, we have
\begin{align}
\dim \ker \matr B_{[n]} = \dim \text{im } \matr B_{[n+1]} \label{eq:equal}
\end{align}
Combining Eqs. \eqref{eq:subset} and \eqref{eq:equal}, we find for every $n>0$
\begin{align}
\text{im } \matr B_{[n+1]} = \ker \matr B_{[n]}.\label{eq:im-eq-ker}
\end{align}

We now apply the rank nullity theorem to $\matr B_{[n]}: \mathbb{R}^{N_{[n]}} \to \mathbb{R}^{N_{[n-1]}}$ which yields
\begin{align}
    \dim \ker \matr B_{[n]} = N_{[n]} - \text{rank } \matr B_{[n]}. \label{eq:ranknullity}
\end{align}
We thus obtain a recurrence relation valid for $n>0$
\begin{align}
  N_{[n]} = \text{rank } \matr B_{[n+1]} + \text{rank } \matr B_{[n]} = \binom{d+1}{n+1} \,,
\end{align}
where we used the fact that for a $d$-simplex, the number of $n$-dimensional faces, formed by choosing $n+1$ nodes from $d+1$ available ones, is $
N_{[n]} = \binom{d+1}{n+1}$. 
Applying the recurrence relation for $n=d,\ldots,1$ together with Pascal's identity then yields Eq.~\eqref{eq:dim-d-n}. As a last step, for $n = 0$, we have to slightly adapt the recurrence relation because 
\begin{align}
    \dim H_0 = 1 = \dim \ker \matr B_{[0]} - \dim \text{im } \matr B_{[1]}.
\end{align}
Nevertheless, this yields the same results as Eq.~\eqref{eq:dim-d-n} for $\vec z^{[+]}$. Finally, there is no $\vec{z}^{[-]}$ for $n=0$ which concludes the proof.

\section{Data availability}
The data generated during the current study can be fully reproduced using the code in the github repository \href{https://github.com/ibfzj/topological-kuramoto/}{https://github.com/ibfzj/topological-kuramoto/}.

\section{Code availability}
The code used to generate the results is openly available at \href{https://github.com/ibfzj/topological-kuramoto/}{https://github.com/ibfzj/topological-kuramoto/}.

\bibliography{references}

\acknowledgements
We thank Ginestra Bianconi for stimulating discussion. D.W. and I.B. gratefully acknowledge support by the German Federal Ministry of Research, Technology and Space (Bundesministerium für Forschung, Technologie und Raumfahrt) via the grant number 03SF0751.
M.T.S acknowledges support by the European Union (ERC, HIGH-HOPeS, 101039827). Views and opinions expressed are however those of the author(s) only and do not necessarily reflect those of the European Union or the European Research Council Executive Agency. Neither the European Union nor the granting authority can be held responsible for them.

\end{document}